\documentclass[%
 aip,
 amsmath,amssymb,twocolumn,
 reprint,%
]{revtex4-1}
\usepackage{graphicx}
\usepackage{dcolumn}

\draft 

\begin{document}
\preprint{AIP/123-QED}

\title{Effects of Rotational Symmetry Breaking in Polymer-coated Nanopores} 



\author{D. Osmanovi\'{c}}
\affiliation{London Centre for Nanotechnology (LCN) and Department of Physics and Astronomy, University College London, WC1E 6BT, London, United Kingdom}
\author{M. Kerr-Winter}
\affiliation{Centre for Mathematics, Physics and Engineering in the Life Sciences and Experimental Biology, University College London, London, WC1E 6BT,  United Kingdom}
\author{R.C. Eccleston}
\affiliation{Centre for Mathematics, Physics and Engineering in the Life Sciences and Experimental Biology, University College London, London, WC1E 6BT,  United Kingdom}
\author{B.W. Hoogenboom}
\affiliation{London Centre for Nanotechnology (LCN) and Department of Physics and Astronomy, University College London, WC1E 6BT, London, United Kingdom}
\author{I.J. Ford}
\affiliation{London Centre for Nanotechnology (LCN) and Department of Physics and Astronomy, University College London, WC1E 6BT, London, United Kingdom}


\date{\today}

\begin{abstract}
The statistical theory of polymers tethered around the inner surface of a cylindrical channel has traditionally employed the assumption that the equilibrium density of the polymers is independent of the azimuthal coordinate. However, simulations have shown that this rotational symmetry can be broken when there are attractive interactions between the polymers. We investigate the phases that emerge in these circumstances, and we quantify the effect of the symmetry assumption on the phase behavior of the system. In the absence of this assumption, one can observe large differences in the equilibrium densities between the rotationally symmetric case and the non-rotationally symmetric case. A simple analytical model is developed that illustrates the driving thermodynamic forces responsible for this symmetry breaking. Our results have implications for the current understanding of the polymer behavior in cylindrical nanopores.
\end{abstract}

\pacs{}

\maketitle 

\section{Introduction}
Nanoscale pores are abundant in biology and are rapidly gaining importance in technological and biomedical applications.\cite{Dekker:2007} Artificial solid-state nanopores enable the detection of single molecules via changes in ionic currents through \cite{Branton:2008,Miles:2013} or across\cite{Ivanov:2011} them. Nanopores can also be used as molecular filtration devices, with applications from separating biomolecules to water purification.\cite{Han:2008} Additionally, nanoporous devices have been used as novel drug delivery devices\cite{Losic:2009}.

Transport selectivity is one of the key challenges in the application of nanopores. While size exclusion is reasonably straightforward, chemical selectivity is harder to achieve. This is exemplified in nanopore-based DNA sequencing. Recent technological advances have improved sensing resolution down to the single nucleotide level,\cite{Cherf:2012,Manrao:2012} but significant challenges remain. In particular, high fidelity and chemically selective sensing depend on an accurate control of the speed with which macromolecules translocate through the nanopore.\cite{Branton:2008,Luan:2012}  Interactions between the DNA and the nanopore surface can lead to significant variation in translocation times,\cite{Branton:2008,Meller:2000,Meller:2001} though well defined distributions have been recorded for shorter DNA fragments.\cite{Li:2010}

 While there is scope for improvement using natural ion channels \cite{Kasianowicz:1996} or by inserting natural protein pores in artificial nanopore devices, \cite{Hall:2010} a more generic method of nanopore functionalization is through the coating of its surface with specific macromolecules. For example, better control of single-molecular transport may be obtained by functionalization of the pores with one-end grafted polymers. \cite{Wanunu:2007,Kowalczyk:2011,Wei:2012,Hou:2011} Solid-state nanopores grafted with polymers are also suggested as a possible flow control mechanism in microfluidic systems, responding to stimuli such as temperature or pH.\cite{Adiga:2005,Adiga:2012}

A fascinating example of a polymer-coated nanopore can be found in the living cell, where the nuclear pore complex (NPC) is responsible for all macromolecular transport between the cell nucleus and the cytoplasm.\cite{Hoelz:2011,Grossman:2012} The walls of the NPC are lined with natively unfolded proteins, which are intrinsically disordered and rich in the phenylalanine-glycine (FG) repeat amino-acid sequence. Macromolecular transport is mediated by untethered globular macromolecules known as nuclear transport receptors via a mechanism that is still disputed, though clearly dependent on the affinity of these receptors to the FG-repeats. NPC transport properties have also been mimicked in artificial devices by coating solid-state nanopores with selected proteins from the NPC central channel.\cite{Jovanovic:2009,Kowalczyk:2011,Kowalczyk:2011:2}

Besides their technological and biological importance, polymer-grafted pores are also of great fundamental interest. In the confinement of a cylindrical nanopore, polymer/polymer interactions have been shown, via numerical simulations,\cite{Peleg:2011,Moussavi:2011,Mincer:2011,Osmanovic:2012,Egorov:2011} to yield a wide and rich pattern of possible morphologies.

The understanding of the possible conformations of polymers within such geometries would be an important advance in the theory of polymer-coated pores and could guide experimental intuition. In particular, whilst work has been done to address the conformations of polymers within nanopores, they almost all use the simplifying assumption that the density of the polymers is rotationally symmetric around the central axis of the nanopore. In this paper, we address the nature of the polymer conformations if this restriction is lifted.


In general, tethered polymer systems have differing phase behavior depending on the intermolecular interactions. These phases are categorized by the scaling behavior of the radius of the gyration of the polymers\cite{Kreer:2004}, and labeled as ``mushrooms" or ``brushes". When the polymers are tethered to the inside of a curved surface, such as the circumference of a nanopore, these behaviors are complemented further by two gross phases that differ from the simple swelling and collapsing of a polymer brush. They can broadly be categorized as ``wall phases" where the majority of the polymer density is found closer to their tethering points, and a ``central phase" where the polymers stretch away, at an entropic cost, from their tethering points and meet in the central channel of the nanopore\cite{Peleg:2011,Osmanovic:2012}. Such behavior could provide a gating mechanism for macromolecular transport through nanopores, which might be exploited in artificial pores, and explain\cite{Osmanovic:2013} the remarkable abilities of the NPC to selectively transport cargos as large as viruses.\cite{Zaitseva:2006}

The primary driver of this change in conformation is the minimization of the surface area of the polymer condensate. However, the wall phase can further reduce its surface area by condensing in separate clumps along the circumference of the pore, such that the average position of the polymers is still close to the wall, but they form into separate regions of large polymer density arranged around the nanopore circumference. Such clumping will have an impact upon the phase diagram of polymer conformations by lowering the free energy of the wall phase.

 In order to understand this better, we determine how rotational symmetry breaking affects the stability of the previously observed central phase\cite{Osmanovic:2012,Peleg:2011}. We do this using several methods: Monte Carlo simulations, density functional theory calculations and a simple analytical treatment.

\section{Methods}
\begin{figure}[tbh]
\begin{center}
\includegraphics[width=\columnwidth]{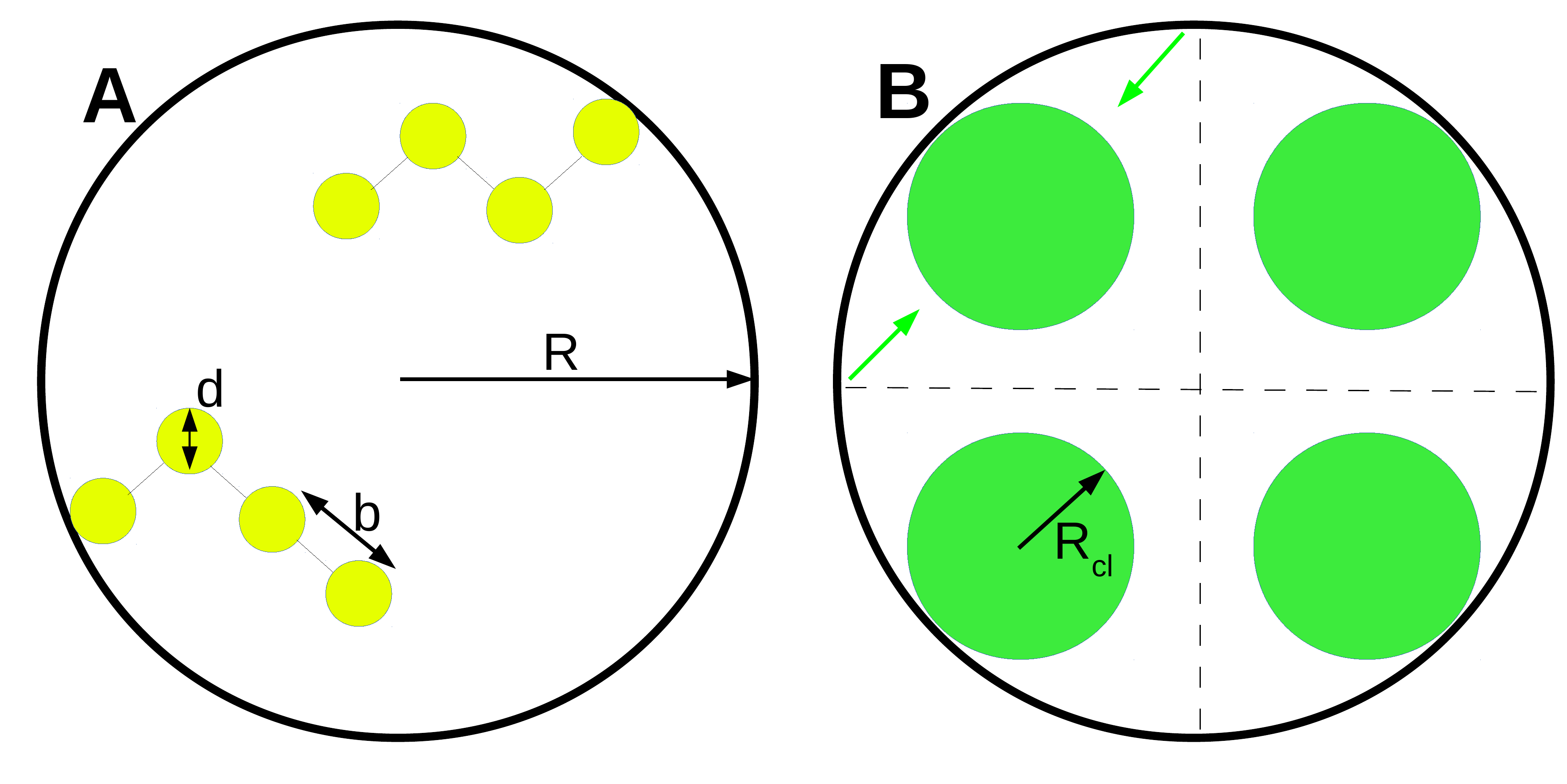}
\caption{A) We consider $M$ polymers of $N$ disks tethered on the inner surface of a circle, illustrated here for the case $M=2$ and $N=4$. B) Basis of the analytical model used to interpret the results. We divide the pore into $N_c$ sections, in each of the tethered polymers stretch to form a clump in the centre of the section, as indicated by the green arrows. In this case $N_c=4$. The change in total free energy is given by the entropic cost of extending the polymers together with the total energy of the polymer phase which depends on the change in the length of the interface it forms with the surrounding medium.}
\label{fig:sch} %
\end{center}
\end{figure}

Our study builds upon on a previous paper \cite{Osmanovic:2012} where we observed rotational symmetry breaking in Monte Carlo simulations of a polymer-coated cylinder. For simplicity and computational tractability we restrict our analysis to two dimensional systems, i.e., here we consider only variations as a function of radial and azimuthal coordinates. The model system is as follows: a number $M$ of polymers consisting of $N$ identical disks tethered around the circumference of a circle of radius $R$, as schematically illustrated in Fig. \ref{fig:sch}A. The disks are separated from each other by a bond length $b$, and the polymers interact with each other through the following potential:
\begin{equation}
\phi(\mathbf{r})=\!\begin{cases}
\infty & |\mathbf{r}|\!<\! d\\
-\epsilon \exp(-(|\mathbf{r}|- d)/\sigma) & |\mathbf{r}|\!\geq\! d\,\end{cases}
\label{eq:ppotential}
\end{equation}
where $d$ is the diameter of the disk, $\mathbf{r}$ is the vector connecting the centers of both beads, $\epsilon$ is the potential depth and $\sigma$ represents the range of the interaction.
The system was studied using Monte Carlo, density functional theory and analytical calculations. The Monte Carlo simulations were performed using the Metropolis algorithm\cite{Metropolis:1953} on systems of tethered polymers and the density functional theory is adapted from our previous work\cite{Osmanovic:2012}. The analytical work is based on simple representations of polymer behavior.

Monte Carlo simulations of 2D polymers with a fixed bond length constraint have difficulties with convergence. If we consider three consecutive disks along a polymer there are only two positions that the central disk of the three can take that preserve the fixed bond length constraint. Because there is only one possible move per disk for any conformation that keeps the bond length constant, Monte Carlo performed in this way is slow to converge. The disk at the end of the polymer is the only one which has a continuous range of possible moves. Moves that rotate a greater number of disks around a fixed disk could be considered, but the difficulty of avoiding disk overlap limits the likelihood that updates which move a large number of disks would be accepted. Therefore the Monte Carlo results presented here should be viewed more as indicative of the equilibrium structure rather than exact quantitative representations of it.

A more robust but necessarily simplified density functional theory of the system was also used to obtain the equilibrium densities of the polymers. This involves constructing a free energy functional of the polymer density and finding the structure that minimizes it. We adapt the functional used previously\cite{Osmanovic:2012} to the here studied 2D system; more details can be found in appendix A.

By considering the change in entropy of the polymers as they stretch away from their tethering points together with the  the total energy of each clump, we can write down a simple form for the change in free energy upon breaking the the rotational symmetry of the wall phase. The total entropy change of the system will be represented as a sum of terms quadratic in the distance between each polymer tether point and the center of its clump. We can evaluate the energy of each clump by integrating the potential Eq. (\ref{eq:ppotential}) over the total area of the clump. This allows us to write the form of the change in free energy against clump number:
\begin{align}
\label{eq:anfe} &\Delta F(N_c)=-N_c \frac{M k_b T }{2 \pi N b^2}\,\,\times \\ \nonumber &\left(\!\!\frac{2 \pi\!  \left( \! R^2\!+\!(R\!-\!R_c)^2 \right) }{N_c}\!+\!4 R(R_c\!-\!R) \sin \left(\!\frac{\pi }{N_c}\!\right)\!\!\right) \\ \nonumber &-\epsilon \exp\left(\!\frac{d}{\sigma}\!\right)\! N M \sigma^2 \pi \rho \exp\left(-\frac{1}{\sigma} \sqrt{\frac{N M}{\pi N_c \rho}} \right)\left( 1\!+\!\frac{1}{\sigma} \sqrt{\frac{N M}{\pi N_c \rho}}\right)
\end{align}
where $N_c$ is the number of clumps, $R_c$ is the radial distance the polymers stretch away from the inner surface of the pore, which is related to the radius of a single clump $R_{cl}$ (see Fig. \ref{fig:sch}B). The only free parameter in this simple analytical model is the polymer disk density $\rho$, and since this parameter varies when we modify $\epsilon$ and $\sigma$, we take this parameter from the theoretical results obtained in density functional theory. Next, the clump number in equilibrium is deteremined by minimizing $\Delta F$ versus $N_c$. Further details about the derivation and assumptions of the model can be found in appendix B.


\section{Results and Discussion}
The calculations were performed with the parameters $M=25$, $N=100$, $b=1.0$ nm, $d=0.25$ nm and $R=25$ nm. The behavior of the system was established for a range of the interaction parameters $\epsilon$ and $\sigma$.

\begin{figure}[tbh]
\begin{center}
\includegraphics[width=\columnwidth]{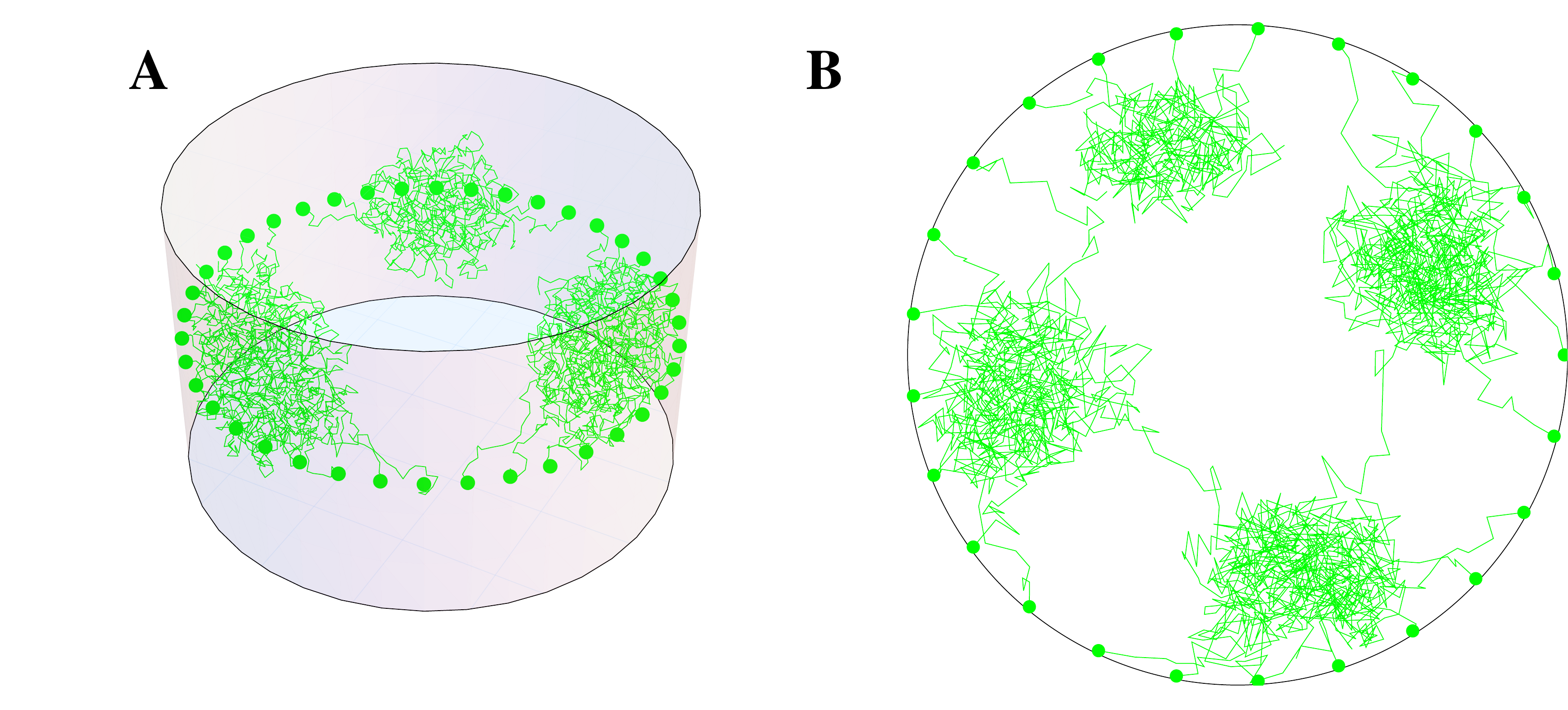}
\caption{Monte Carlo simulations of polymers tethered on a circular ring on the inside of a 3D cylinder (A) and to the inside of a circle in 2D (B). The Monte Carlo simulations show that clumping behavior can emerge in both two and three dimensions for interaction parameters  (A) $\epsilon =  0.1 k_b T,
\sigma = 1.00$ nm (B)$\epsilon =  0.04 k_b T, \sigma = 1.25 nm$, motivating more quantitative methods for determining the clump phase. }
\label{fig:mc1} %
\end{center}
\end{figure}

Monte Carlo simulations can illustrate the propensity of the system to display rotationally asymmetric behavior. Fig. \ref{fig:mc1} gives an example of the clumping behavior of polymers in a pore.\cite{Osmanovic:2012}
The system breaks rotational symmetry for certain parameters in both two and three dimensions.

Convergence issues with Monte Carlo, discussed in section III, potentially limit the reliability of this approach. For a more accurate investigation, we adapt our previously published density functional theory\cite{Osmanovic:2012} to describe two-dimensional polymers within a circular pore. Density functional theory can typically provide results with less computational effort than Monte Carlo. In addition, free energies can be obtained directly. We perform calculations with and without the rotational symmetry constraint. By comparing the resulting phase diagrams, the effect of rotational symmetry breaking on the phase of the system can be determined.
\begin{figure}[tbh]
\begin{center}
\includegraphics[width=\columnwidth]{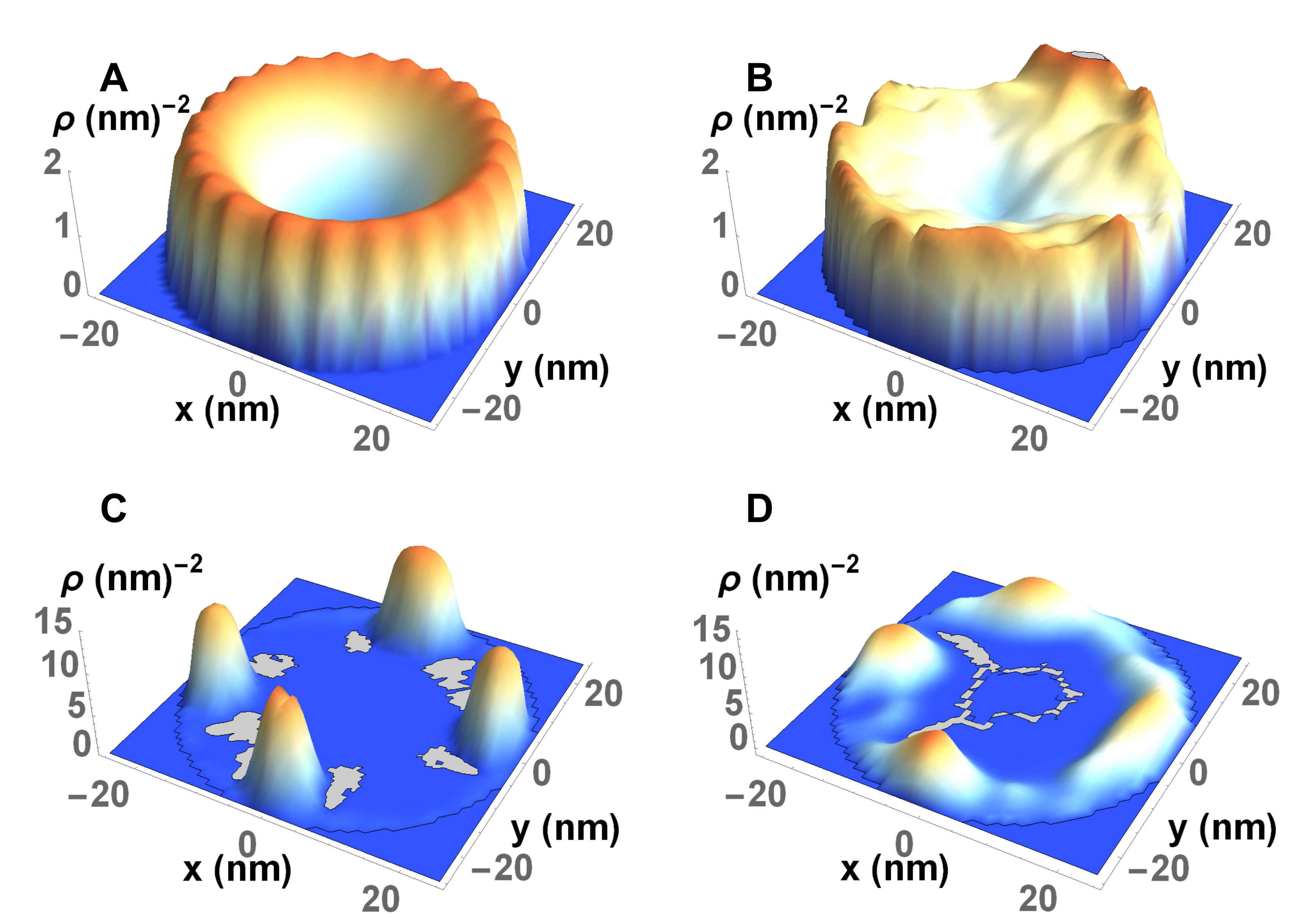}
\caption{Equilibrium densities of polymer disks for density functional theory and Monte Carlo, for $\epsilon=0.01 \mathrm{k}_b T$ and $\sigma=0.75$ nm (A and B respectively) and for $\epsilon=0.04 \mathrm{k}_b T$ and $\sigma=1.25$ nm (C and D respectively). Both methods yield similar results. }
\label{fig:dftmc} %
\end{center}
\end{figure}
We verify density functional theory results for the predicted clump number by Monte Carlo simulations for the same parameters.

 Because of the problems in the convergence of Monte Carlo, precise quantitative agreement between density functional theory and Monte Carlo is hard to achieve. Nevertheless, there is a qualitative similarity in the obtained results from the two different methods, as can be seen from Fig. \ref{fig:dftmc}.

\begin{figure*}
\begin{center}
\includegraphics[width=\textwidth]{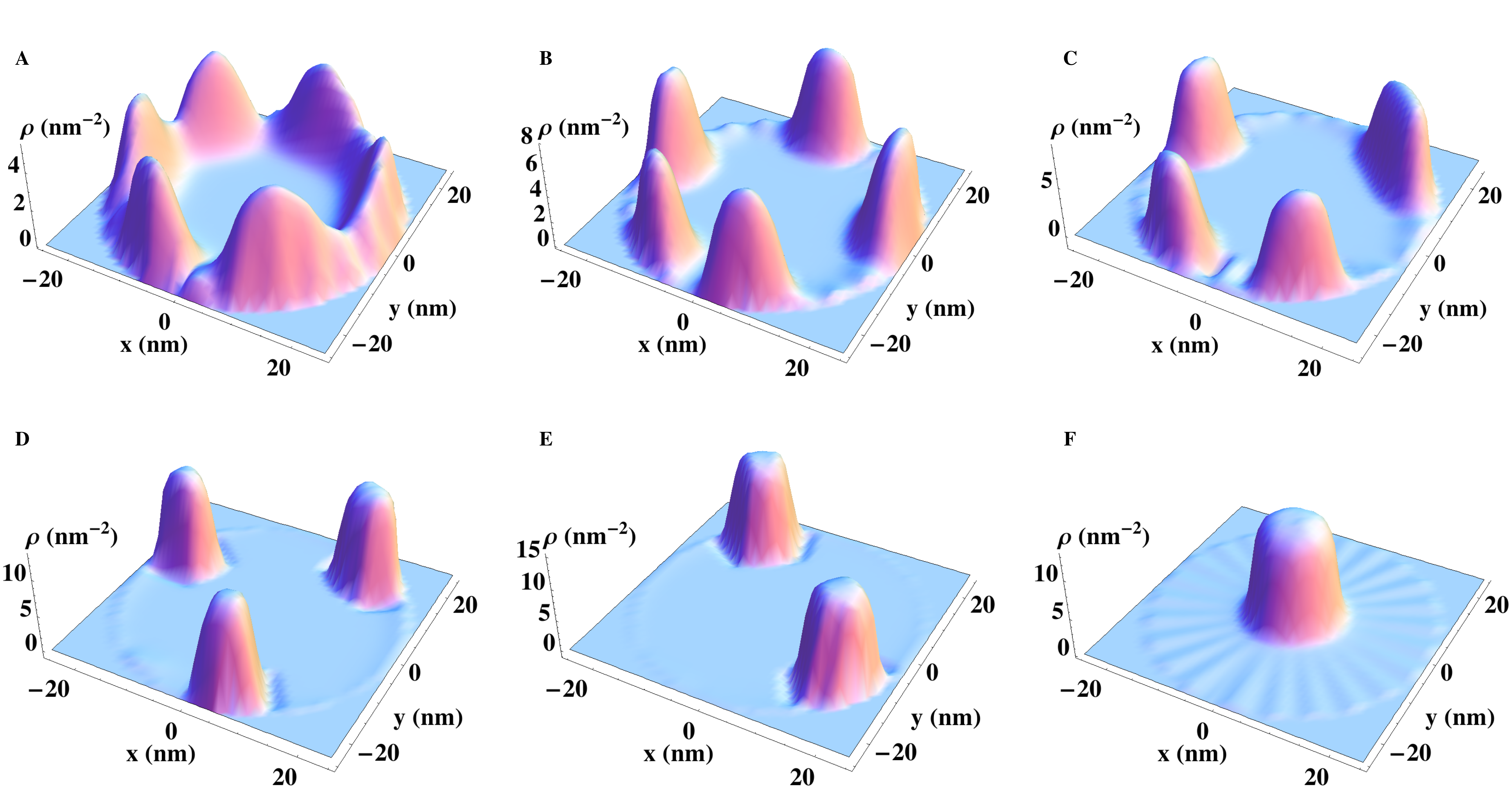}
\caption{Density in a circle of radius $R=25$ of 2D constrained polymers with parameters $M=25, b=1 \text{nm}, N=100,d=0.25 \text{nm}$ and various $\epsilon$ and $\sigma$ (A) $\epsilon=0.03 k_b T,\sigma=0.75$ nm, (B) $\epsilon=0.04 k_b T,\sigma=0.75$ nm, (C) $\epsilon=0.05 k_b T,\sigma=0.75$ nm, (D) $\epsilon=0.06 k_b T,\sigma=1.25$ nm, (E) $\epsilon=0.07 k_b T,\sigma=1.75$ nm, (F) $\epsilon=0.05 k_b T,\sigma=1.25$ nm, obtained from density functional theory. There exist both the central phase (F) and various wall phases (A-E) with different numbers of clumps.   }
\label{fig:den}
\end{center}
\end{figure*}

Fig. \ref{fig:den} shows an example of the various phases obtained from the density functional theory approach. It is observed that there is a multiplicity of wall states, with different numbers of clumps. There exist many metastable states for each parameter set, necessitating the calculations to be repeated many times to determine the phase with the lowest free energy. The central and the wall phase can both exist at the same parameter set.

The wall phase can consist of any number of clumps up to the number of polymers in the system. Despite the differing numbers of clumps, it appears that the position of the clumps reflect maximal symmetry, in the sense that when there are six clumps they appear to be positioned on the vertices of a hexagon; when there are five clumps they lie on the vertices of a pentagon and so forth. Additionally, all the clumps are similar in terms of disk density and spatial extent.


According to the analytical model described in appendix B, the total entropy cost is, very approximately, inversely proportional to the clump number. Therefore, large numbers of clumps carry little entropic cost of formation compared to a single central clump. Again approximately, the total energy cost increases logarithmically with clump number (up to an asymptote). The clumping number is determined by the delicate balance between these energetic and entropic contributions to the free energy.

Using the density functional theory method we can test whether the rotational symmetry assumption has a large impact on the threshold interaction parameters at which the system has to choose between central and wall states. We can then investigate how well the simple analytical model can describe what is occuring in the phase transition.

\begin{figure*}[t!]
\begin{center}
\includegraphics[width=\textwidth]{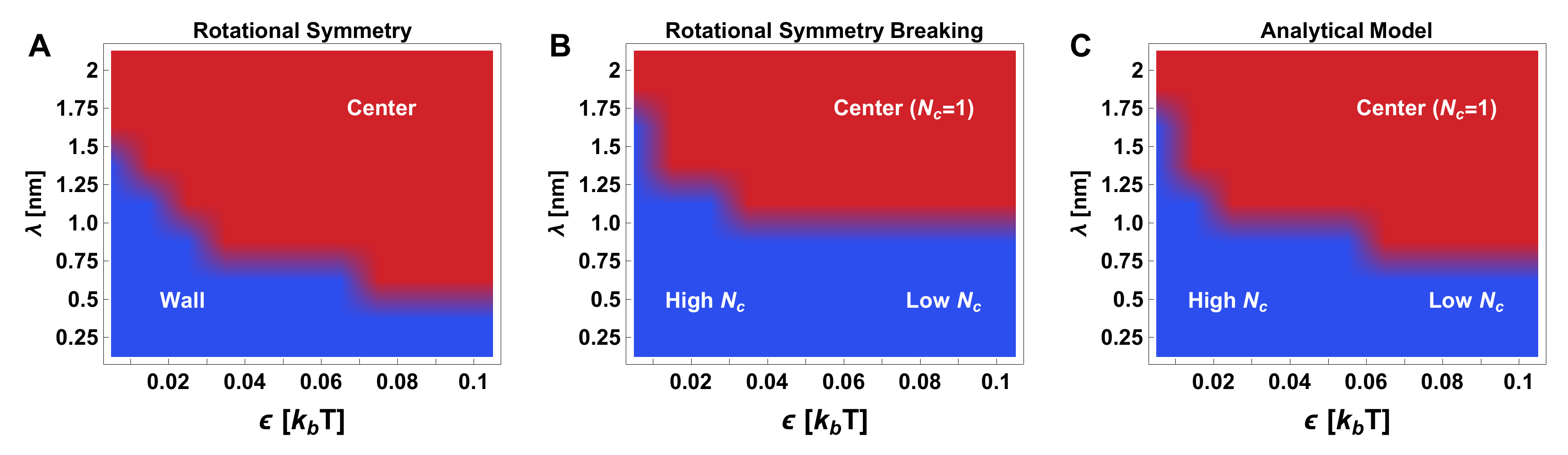}
\caption{Phase diagram for 2D polymers with A) imposed rotational symmetry assumption, B) broken rotational symmetry and C) the analytical model. The blue regions are phases where the majority of the polymer density lies near the wall and the red regions are where the majority of the polymer density can be found near the central region.}
\label{fig:phasemap} %
\end{center}
\end{figure*}

Fig. \ref{fig:phasemap} shows the phase diagram of the polymers in terms of central/wall phases for the cases with the rotational symmetry assumption (A) and without it (B). We can see clearly that relaxing this assumption leads to a wall phase that persists over a larger range $\sigma$ and $\epsilon$. However, the overall effect is not too large, indeed a shift of around $0.25$ nm in $\sigma$ would lead to phase diagrams that are very similar. The analytical model (C) captures the behavior, despite its many simplifications.

The success of the analytical model confirms that the clumping behavior results from a competition between the entropy of stretching the polymers against the energy penalty of creating a larger surface area. When there are large numbers of clumps, the polymers do not need to stretch very far, as the clump is likely to be close to the polymer tethering point, and therefore the entropic cost of forming more numerous clumps is lower. However, a larger number of clumps will have a greater surface area compared to fewer clumps, which will increase the energy penalty. This competition determines the equilibrium phase, and as we decrease the strength of the attraction the system will begin to favor states with more clumps. The analytical model gives a simple way of quantifying, for a particular set of parameters, how this behaviour emerges.

\section{Conclusion}
We have shown that a system of polymers tethered around the circumference of a nanopore can break rotational symmetry, leading to the formation of clumps at the wall. The number of clumps can be predicted from the system parameters, such as the interaction strength and the radius of the pore. Previous theoretical results showed that tethered polymers could adopt wall or central phases, and this phase description remains valid even if the rotational symmetry assumption is dropped. When the rotational symmetry assumption is relaxed, the wall phase divides into multiple clumps which leads to a slight lowering of the free energy of the wall phase compared to the symmetric case, shrinking the boundaries of central phase stability on the phase diagram.

This effect will be largely dependent on the parameters chosen. For instance, cylindrical pores of small radii with long polymers will likely always be completely clogged by polymer, regardless of rotational symmetry breaking. This is similar to what is to be expected in three dimensions, such that previous results obtained in three dimensions are not qualitatively altered.

We have developed a simple analytic model of the system that compares favorably with density functional theory calculations. By considering only the simplest form contributions to the free energy might take, we derived an equation that gives the change in free energy as the number of clumps is varied. This simple form of the free energy can serve to guide intution.

\section{Acknowledgements}
The authors acknowledge the use of the UCL Legion High Performance Computing Facility (Legion@UCL), and associated support services, in the completion of this work. This work was funded by the Sackler Foundation and by the UK Engineering and Physical Sciences Research Council (EP/F500351/1).

\appendix

\section{Density Functional Theory}
For density functional theory calculations we use the following free energy functional, slightly modified from previous work\cite{Osmanovic:2012} for the system considered.
\begin{align}
\beta F[W(\mathbf{r})] & =-\sum_{r_0}\ln Z[W(\mathbf{r}),\mathbf{r_0}]-\int W(\mathbf{r})\rho(\mathbf{r})\,\mathrm{d}\mathbf{r}\nonumber\\
 & \nonumber +\int V(\mathbf{r})\rho(\mathbf{r})\mathrm{d}\mathbf{r} +\int\mathrm{d}\mathbf{r}\,\left\{ \Phi^{HD}(\mathbf{r})+\Phi^{ch}(\mathbf{r})\right\}\\
 & +\frac{1}{2}\iint \mathrm{d}\mathbf{r}\mathrm{d}\mathbf{r'} \rho(\mathbf{r})\rho(\mathbf{r'})\phi(\mathbf{r}-\mathbf{r'})
\label{eq:polymer_func}
\end{align}
where $F$ is the total free energy of the system. $W(\mathbf{r})$ is a ``mean-field" potential, which can be mapped to a disk density $\rho(\mathbf{r})$. $V(\mathbf{r})$ is the external potential that acts on the polymers, here set to 0. $\Phi^{HD}$ gives the excess free energy density arising from the excluded volume imposed by the hard disks, and is given by\cite{Mulero:20082}
\begin{equation}
\Phi^{HD}(\mathbf{r}) = \rho(\mathbf{r})\left( \frac{\eta(\mathbf{r})}{1-\eta(\mathbf{r})}-\ln(1-\eta(\mathbf{r}))\right)\,\,\, \text{,}
\end{equation}
where $\eta(\mathbf{r})$ is the packing fraction in two dimensions, given by $\rho(\mathbf{r})\pi d^2/4$ with $d$ the diameter of the disk. $\Phi^{ch}$ is the ``chain connectivity" term, describing excess free energy due to the fact the hard disks are joined together. It is given by\cite{Wertheim:1987,Egorov:2008}
\begin{equation}
\Phi^{ch}(\mathbf{r}) =\frac{1-N}{N}\ln\left(\frac{2-\eta(\mathbf{r})}{2(1-\eta(\mathbf{r}))^3}\right)
\end{equation}

We introduce a Green's function $G(\mathbf{r}_{0},\mathbf{r},s;[W])$ that satisfies
\begin{equation}
\frac{\partial G(\mathbf{r}_{0},\mathbf{r},s;[W])}{\partial s}=\left(\frac{b^{2}}{4}\nabla^{2}-W(\mathbf{r})\right)G(\mathbf{r}_{0},\mathbf{r},s;[W])
\label{eq:green}
\end{equation}
where $b$ is the segment length of the polymer. The Green's function describes the probability that a polymer of $s$ segments will extend from $\mathbf{r_0}$ to $\mathbf{r}$ in the presence of the potential $W$. This equation is to be solved for every tethering point $\mathbf{r_0}$ of the polymers around the circumference of the circle. From $G(\mathbf{r}_{0},\mathbf{r},s;[W])$ we calculate the density in the following way:
\begin{equation}
\rho(\mathbf{r})=\frac{\int_{0}^{N}\mathrm{d}s\int\mathrm{d}\mathbf{r^{\prime}}\sum_{r_{0}}G(\mathbf{r}_{0},\mathbf{r},N-s;[W])G(\mathbf{r},\mathbf{r^{\prime}},s;[W])}{\int\mathrm{d}\mathbf{r^{\prime}}\sum_{r_{0}}G(\mathbf{r}_{0},\mathbf{r^{\prime}},N;[W])}
\label{eq:grho}
\end{equation}
thereby taking into account polymers attached to every tethering point $\mathbf{r_0}$ in the system. $Z[W(\mathbf{r}),\mathbf{r_0}]$ represents the partition function of a polymer tethered at $\mathbf{r_0}$ and interacting with the mean field. It is given by:
\begin{equation}
Z[W(\mathbf{r}),\mathbf{r_0}]=\int G(\mathbf{r}_{0},\mathbf{r},N;[W])\mathrm{d}\mathbf{r} \,\,\,\, \text{.}
\end{equation}

The method for obtaining equilibrium densities requires the minimization of the free energy $F[W(\mathbf{r})]$. We use a steepest descent scheme involving the functional derivative of $F$:
\begin{equation}
\frac{\partial W(\mathbf{y})}{\partial t}=\frac{\delta F[W(\mathbf{r})]}{\delta W(\mathbf{y})}
\end{equation}
where we have introduced a fictional time variable $t$. Discretizing this equation leaves us with the following form, which is useful for calculations:
\begin{equation}
W^{n+1}(\mathbf{y})=W^n(\mathbf{y}) + \Delta t \frac{\delta F[W(\mathbf{r})]}{\delta W^n(\mathbf{y})}
\label{eq:rule}
\end{equation}

The complete method for solving for equilibrium densities involves the following: the mean field is initialized to a certain profile $W_0$, from which Eq. (\ref{eq:green}) and Eq. (\ref{eq:grho}) are used to obtain the density $\rho$. Once the density is obtained, the functional derivative of the free energy at this mean field and density is calculated, and this is then used to update the mean field according to equation (\ref{eq:rule}). This process is repeated until the system has reached convergence, i.e. the free energy and density no longer change after updates. This process is repeated ten times for each parameter set in order to find the global minimum for each case. It is important to note that the field $W_0$ is initialised to random values using a random number generator. This ensures that the rotational symmetry is broken: if this were not imposed, the algorithm would maintain the symmetry, as in our previous study\cite{Osmanovic:2012}.

\section{Analytical Model}

We denote the number of clumps as $N_c$ and we make the following assumptions about them and the rest of the system:
\begin{itemize}
\item The clumps are circular with a radius $R_{cl}$. If the clumps lie against the wall their centers will be a distance $R_{cl}$ from the wall. If there is only one clump it will lie in the center of the circle. The clumps are identical. We represent the distance of the center of a clump from the wall of the pore  as $R_c$, where $R_c$ depends on $N_c$ and is given by:
\begin{equation}
R_c=\!\begin{cases}
R & N_c=1\\
R_{cl} & N_c > 1\end{cases}
\label{eq:rc}
\end{equation}
\item The system maintains ``maximal symmetry" for instance, if there are 3 clumps, a rotation of the system by $\frac{2\pi}{3}$ will leave the system invariant. In other words, the clumps will lie on the vertices of a regular polygon. For practical purpose we divide the circle into $N_c$ equal sectors, and all polymers tethered within one section form one clump at the center of that section.
\item The density of disks within a clump is constant, and is related to the inverse of the disk diameter squared.
\item The entropy difference between states is given by the specific entropic costs of extending the polymers from their tether points to the center of each clump.

\item The entropic contribution to the free energy of clump formation from a uniform wall configuration will be represented by a sum of terms of the form:
\begin{equation}
\Delta S_1(\mathbf{r})=-\frac{k_b}{N b^2}r^2
\end{equation}
for each polymer where $r$ is the end to end distance, $k_b$ is the Boltzmann constant.
\end{itemize}

The system being treated is shown in Fig. \ref{fig:sch}. In this case, $N_c=4$. The polymers stretch from each quadrant edge towards the center of that quadrant's clump.

The free energy change of this system is given by:
\begin{equation}
\Delta F= \Delta E - T \Delta S
\end{equation}
Where $T$ is the temperature of the polymer.
\begin{figure}[tbh]
\begin{center}
\includegraphics[width=\columnwidth]{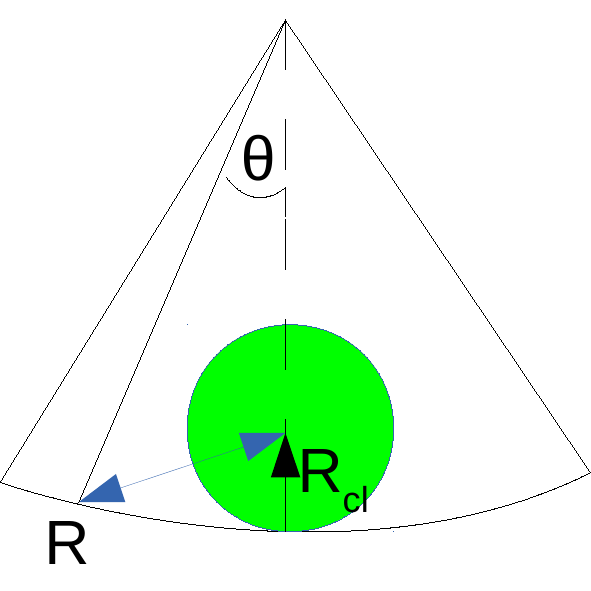}
\caption{The geometry used to calculate the entropy of extension of the polymers. The polymers tethered around the edge of the pore stretch to the center of the clump, at $\theta$ = 0 and $r = R-R_{cl}$. In order to calculate the entropy change, contributions proportional to the square of this extension are summed over polymer tether points lying on the edge of the circle.}
\label{fig:analsch}
\end{center}
\end{figure}
The total entropy change $\Delta S$ will be given by the entropic cost of extending the polymers:
\begin{equation}
T \Delta S=-N_c \frac{k_b T}{N b^2}\!\!\sum_{i=1}^{M/N_c}\!\!\! \left(R^2\!+\!(R\!-\!R_c)^2\!-\!2 (R\!-\!R_c) R \cos(\theta_i)\right)\,\,.
\end{equation}
This can be understood as a summation of the contributions of all the polymers in a sector as they stretch from $r=R$ and $\theta=\theta_i$ to $r=R-R_{c}$ and $\theta=0$, where $\theta$ is the angle along the sector. This is illustrated in Fig. \ref{fig:analsch}. For analytical reasons, we assume that the polymers are tethered uniformly and continuously along the circumference of the circle, in which case we would modify our discrete representation to:
\begin{align}
T\Delta S(N_c)&=-N_c \frac{k_b T}{N b^2} \rho_s  \times \nonumber \\ &\int_{-\pi/N_c}^{\pi/N_c}\!\!\!\!\!\! \left(\!R^2\!+\!(R\!-\!R_c)^2\!-\!2 (R\!-\!R_c) R \cos(\theta)\right)\mathrm{d}\theta
\end{align}
where $\rho_s$ is the number of polymers tethered per unit length of pore circumference, i.e $2\pi\rho_s=M$

This integral can be evaluated to yield:
\begin{align}
T\Delta S(N_c)&=-N_c \frac{M k_b T }{2 \pi N b^2}  \times \nonumber \\ &\left(\!\!\frac{2 \pi  \left( \! R^2\!+\!(R\!-\!R_c)^2 \right) }{N_c}\!+\!4 R(R_c\!-\!R) \sin \left(\frac{\pi }{N_c}\right)\!\!\right)
\end{align}
For example when $N_c = 1$ we use the definition of $R_c$ given before to give the total entropy change as:
\begin{equation}
T \Delta S(N_c=1)=-\frac{ M k_b T R^2}{ N b^2}
\end{equation}

We use the incompressibility condition to calculate the clump radius:
\begin{equation}
\frac{N M}{N_c \pi R_{cl}^2}=\frac{4 \eta}{\pi d^2}
\end{equation}
such that
\begin{equation}
R_{cl} = \sqrt{\frac{N M}{4 N_c \eta} } d
\end{equation}
Where $\eta$ is the packing fraction, as defined in density functional theory.




The total internal energy of a clump is given by the sum of interdisk interactions within the clump. We assume that the inter-clump interactions are negligible. This assumption will hold when there are few clumps, such that the distance between clumps is large.
\begin{equation}
E_c=\frac{1}{2}\int \rho(\mathbf{r'})\rho(\mathbf{r})\phi(\mathbf{r}-\mathbf{r'})\mathrm{d}\mathbf{r}\mathrm{d}\mathbf{r'}
\end{equation}
To calculate the internal energy, we integrate over the potential in a clump using a radial coordinate system.

The full form of the interdisk interaction is:
\begin{equation}
\phi(\mathbf{r}-\mathbf{r'})=\epsilon \exp\left(-\frac{|\mathbf{r}\!-\!\mathbf{r'}|\!-\!d}{\sigma}\right)\Theta\left(|\mathbf{r}\!-\!\mathbf{r'}|\!-\!d\right)
\end{equation}
where $\Theta$ is the Heaviside step function and $|\mathbf{r}\!-\!\mathbf{r'}|$ is given by (in plane polar coordinates):
\begin{equation}
|\mathbf{r}\!-\!\mathbf{r'}|=(r^2+r'^2-2 r r' \cos(q-q'))^\frac{1}{2}
\end{equation}
where $r$ and $r'$ are the radial positions, relative to the centre of the clump, of points $\mathbf{r}$ and $\mathbf{r'}$ and $q$ and $q'$ are the angular coordinates with respect to the clump center.
 Using the assumption that the disk density in the clump is constant and rearranging we obtain:
\begin{align}
E_c&=\epsilon\rho^2 \pi \exp\left(\frac{d}{\sigma}\right)\!\!\! \int_0^{R_{cl}} \!\!\!\!\!\!\mathrm{d}r\!\!\int_0^{R_{cl}}\!\!\!\!\!\!\!\mathrm{d}r' \!\!\!\int_0^{2\pi} \!\!\!\!\!\!\mathrm{d}q r r'\exp\left(-\frac{|\mathbf{r}\!-\!\mathbf{r'}|}{\sigma}\right)\\ \nonumber&\times\Theta\left(|\mathbf{r}\!-\!\mathbf{r'}|\!-\!d\right)
\end{align}
where we have performed the integration over one of the angles. If we assume that the diameter of the bead is small in comparison to the radius of the clump (again true for low numbers of clumps) then we ignore the Heaviside function for analytical tractability, giving:
\begin{equation}
E_c=\pi \epsilon \rho^2 \exp\left(\frac{d}{\sigma}\right)R_c^4\!\!\! \int_0^{1} \!\!\!\!\mathrm{d}x\!\!\int_0^{1}\!\!\!\!\!\mathrm{d}x' \!\!\!\int_0^{2\pi} \!\!\!\!\!\!\mathrm{d}q x x'\exp\left(-\frac{R_c}{\sigma}|\mathbf{x}\!-\!\mathbf{x'}|\right)
\end{equation}
where $x=r/R_c$ and $x'=r'/R_c$. The integral can be evaluated approximately to give:
\begin{equation}
E_c=\epsilon\rho^2 \exp\left(\frac{d}{\sigma}\right) \pi^2 R_c^2 \sigma^2\left(1\!-\!\exp\left(-\frac{R_c}{\sigma}\right)\left(1\!+\!\frac{R_c}{\sigma}\right)\right)
\end{equation}
and the total energy will be given by $E=N_c E_c$. Using the relation between $R_c$ and $N_c$ allows us to write the energy in terms of clump number.
\begin{align}
E=&\epsilon \exp\left(\frac{d}{\sigma}\right) N M \sigma^2 \pi \rho \,\, \times \\ \nonumber &\left(1-\exp\left(-\frac{1}{\sigma} \sqrt{\frac{N M}{\pi N_c \rho}} \right)\left( 1+\frac{1}{\sigma} \sqrt{\frac{N M}{\pi N_c \rho}}\right)\right)
\end{align}
such that the energy change as a function of the number of clumps is given by:
\begin{align}
\Delta E &= -\epsilon \exp\left(\frac{d}{\sigma}\right) N M \sigma^2 \pi \rho\,\,\times \\&\nonumber\exp\left(-\frac{1}{\sigma} \sqrt{\frac{N M}{\pi N_c \rho}} \right)\left( 1+\frac{1}{\sigma} \sqrt{\frac{N M}{\pi N_c \rho}}\right)
\end{align}

\begin{figure}[tbh]
\begin{center}
\includegraphics[width=\columnwidth]{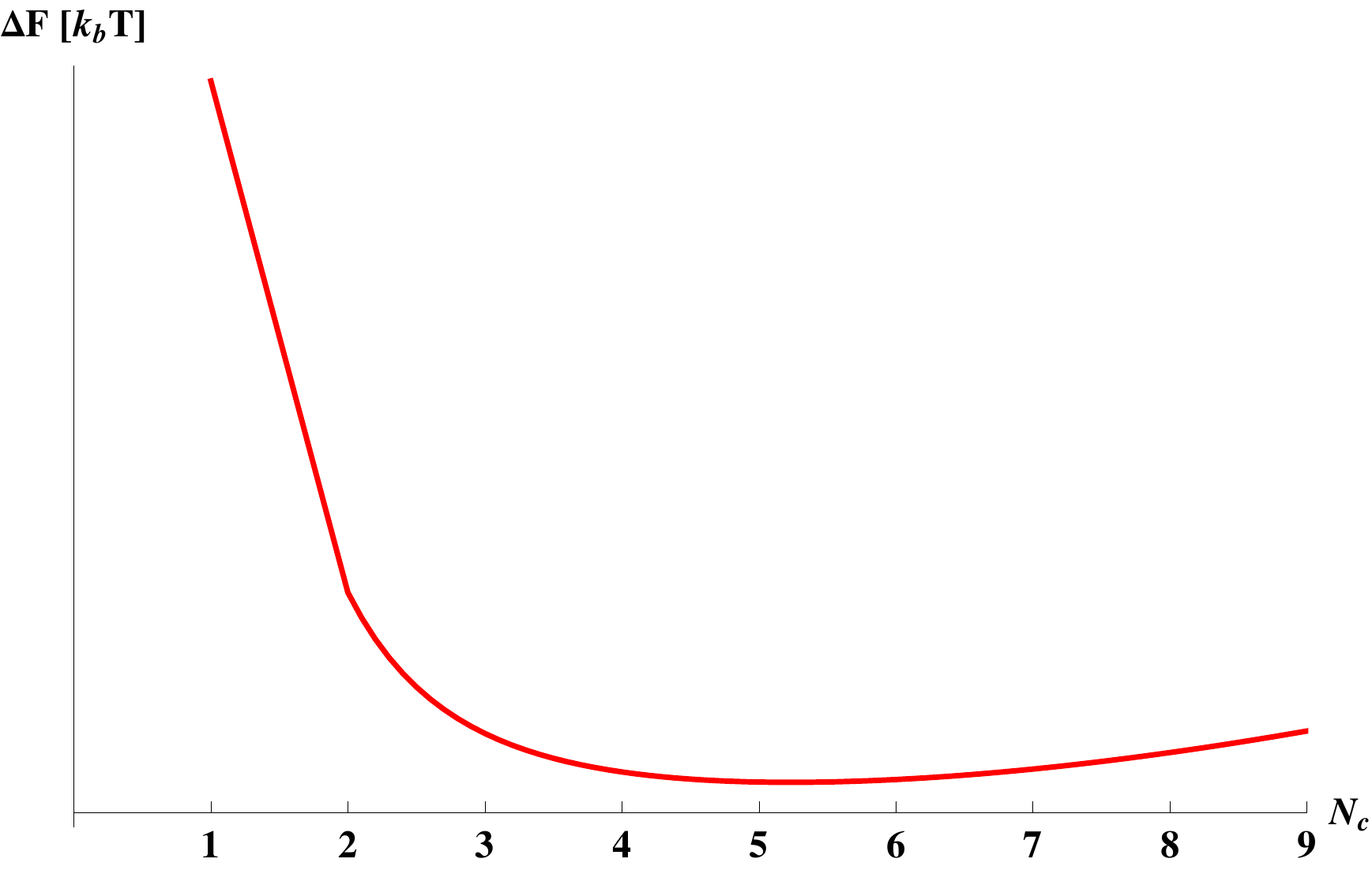}
\caption{Form of the analytic free energy change against the clump number $N_c$ for a specific set of parameters: $N = 100, \, M = 25,\, d = 0.25\, \text{nm},\, R = 25 \, \text{nm} ,
  \, b = 1 \, \text{nm}, \epsilon = 0.07 \,\text{k}_b T,\, \sigma = 0.5\, \text{nm}$ with these parameter sets the packing fraction $\eta$ is $0.383989$.}
\label{fig:fean}
\end{center}
\end{figure}

The combined entropy and energy are given in Eq. (\ref{eq:anfe}) in the main text. An example of the free energy change against clump number is shown in Fig. \ref{fig:fean}.

Using similar arguments, we are able to derive a form of the free energy for the equivalent system in three dimensions:
\begin{align}
\Delta F_{3D} &= \frac{3 M k_b T \left(\frac{2 \pi  \left(\left(R-R_c\right){}^2+R^2\right)}{N_c}+4 R \left(R_c-R\right) \sin \left(\frac{\pi
   }{N_c}\right)\right)}{4 \pi  N b^2 }\\ \nonumber &-\frac{12 \eta  M N \sigma ^3 \epsilon \, e^{d/\sigma } \left(2-e^{-\frac{R_c}{\sigma }}
   \left(\frac{R_c^2}{\sigma ^2}+\frac{R_c}{\sigma }+2\right)\right)}{d^3}
\end{align}
with $R_c=\frac{1}{2} d \sqrt[3]{\frac{M N}{\eta  N_c}}$ when $N_c \ge 2$ and $R_c=R$ when $N_c=1$.


\begin{thebibliography}{36}%
\makeatletter
\providecommand \@ifxundefined [1]{%
 \@ifx{#1\undefined}
}%
\providecommand \@ifnum [1]{%
 \ifnum #1\expandafter \@firstoftwo
 \else \expandafter \@secondoftwo
 \fi
}%
\providecommand \@ifx [1]{%
 \ifx #1\expandafter \@firstoftwo
 \else \expandafter \@secondoftwo
 \fi
}%
\providecommand \natexlab [1]{#1}%
\providecommand \enquote  [1]{``#1''}%
\providecommand \bibnamefont  [1]{#1}%
\providecommand \bibfnamefont [1]{#1}%
\providecommand \citenamefont [1]{#1}%
\providecommand \href@noop [0]{\@secondoftwo}%
\providecommand \href [0]{\begingroup \@sanitize@url \@href}%
\providecommand \@href[1]{\@@startlink{#1}\@@href}%
\providecommand \@@href[1]{\endgroup#1\@@endlink}%
\providecommand \@sanitize@url [0]{\catcode `\\12\catcode `\$12\catcode
  `\&12\catcode `\#12\catcode `\^12\catcode `\_12\catcode `\%12\relax}%
\providecommand \@@startlink[1]{}%
\providecommand \@@endlink[0]{}%
\providecommand \url  [0]{\begingroup\@sanitize@url \@url }%
\providecommand \@url [1]{\endgroup\@href {#1}{\urlprefix }}%
\providecommand \urlprefix  [0]{URL }%
\providecommand \Eprint [0]{\href }%
\providecommand \doibase [0]{http://dx.doi.org/}%
\providecommand \selectlanguage [0]{\@gobble}%
\providecommand \bibinfo  [0]{\@secondoftwo}%
\providecommand \bibfield  [0]{\@secondoftwo}%
\providecommand \translation [1]{[#1]}%
\providecommand \BibitemOpen [0]{}%
\providecommand \bibitemStop [0]{}%
\providecommand \bibitemNoStop [0]{.\EOS\space}%
\providecommand \EOS [0]{\spacefactor3000\relax}%
\providecommand \BibitemShut  [1]{\csname bibitem#1\endcsname}%
\let\auto@bib@innerbib\@empty
\bibitem [{\citenamefont {Dekker}(2007)}]{Dekker:2007}%
  \BibitemOpen
  \bibfield  {author} {\bibinfo {author} {\bibfnamefont {C.}~\bibnamefont
  {Dekker}},\ }\href {\doibase 10.1038/nnano.2007.27} {\bibfield  {journal}
  {\bibinfo  {journal} {Nat. Nanotechnol.}\ }\textbf {\bibinfo {volume} {2}},\
  \bibinfo {pages} {209} (\bibinfo {year} {2007})}\BibitemShut {NoStop}%
\bibitem [{\citenamefont {Branton}\ \emph {et~al.}(2008)\citenamefont
  {Branton}, \citenamefont {Deamer}, \citenamefont {Marziali}, \citenamefont
  {Bayley}, \citenamefont {Benner}, \citenamefont {Butler}, \citenamefont
  {Di~Ventra}, \citenamefont {Hibbs}, \citenamefont {Huang},\ and\
  \citenamefont {Jovanovich}}]{Branton:2008}%
  \BibitemOpen
  \bibfield  {author} {\bibinfo {author} {\bibfnamefont {D.}~\bibnamefont
  {Branton}}, \bibinfo {author} {\bibfnamefont {D.~W.}\ \bibnamefont {Deamer}},
  \bibinfo {author} {\bibfnamefont {A.}~\bibnamefont {Marziali}}, \bibinfo
  {author} {\bibfnamefont {H.}~\bibnamefont {Bayley}}, \bibinfo {author}
  {\bibfnamefont {S.~A.}\ \bibnamefont {Benner}}, \bibinfo {author}
  {\bibfnamefont {T.}~\bibnamefont {Butler}}, \bibinfo {author} {\bibfnamefont
  {M.}~\bibnamefont {Di~Ventra}}, \bibinfo {author} {\bibfnamefont
  {A.}~\bibnamefont {Hibbs}}, \bibinfo {author} {\bibfnamefont
  {X.}~\bibnamefont {Huang}}, \ and\ \bibinfo {author} {\bibfnamefont {S.~B.}\
  \bibnamefont {Jovanovich}},\ }\href@noop {} {\bibfield  {journal} {\bibinfo
  {journal} {Nat. Biotechnol.}\ }\textbf {\bibinfo {volume} {26}},\ \bibinfo
  {pages} {1146} (\bibinfo {year} {2008})}\BibitemShut {NoStop}%
\bibitem [{\citenamefont {Miles}\ \emph {et~al.}(2013)\citenamefont {Miles},
  \citenamefont {Ivanov}, \citenamefont {Wilson}, \citenamefont {Dogan},
  \citenamefont {Japrung},\ and\ \citenamefont {Edel}}]{Miles:2013}%
  \BibitemOpen
  \bibfield  {author} {\bibinfo {author} {\bibfnamefont {B.~N.}\ \bibnamefont
  {Miles}}, \bibinfo {author} {\bibfnamefont {A.~P.}\ \bibnamefont {Ivanov}},
  \bibinfo {author} {\bibfnamefont {K.~A.}\ \bibnamefont {Wilson}}, \bibinfo
  {author} {\bibfnamefont {F.}~\bibnamefont {Dogan}}, \bibinfo {author}
  {\bibfnamefont {D.}~\bibnamefont {Japrung}}, \ and\ \bibinfo {author}
  {\bibfnamefont {J.~B.}\ \bibnamefont {Edel}},\ }\href {\doibase
  10.1039/C2CS35286A} {\bibfield  {journal} {\bibinfo  {journal} {Chem. Soc.
  Rev.}\ }\textbf {\bibinfo {volume} {42}},\ \bibinfo {pages} {15} (\bibinfo
  {year} {2013})}\BibitemShut {NoStop}%
\bibitem [{\citenamefont {Ivanov}\ \emph {et~al.}(2011)\citenamefont {Ivanov},
  \citenamefont {Instuli}, \citenamefont {McGilvery}, \citenamefont {Baldwin},
  \citenamefont {McComb}, \citenamefont {Albrecht},\ and\ \citenamefont
  {Edel}}]{Ivanov:2011}%
  \BibitemOpen
  \bibfield  {author} {\bibinfo {author} {\bibfnamefont {A.~P.}\ \bibnamefont
  {Ivanov}}, \bibinfo {author} {\bibfnamefont {E.}~\bibnamefont {Instuli}},
  \bibinfo {author} {\bibfnamefont {C.~M.}\ \bibnamefont {McGilvery}}, \bibinfo
  {author} {\bibfnamefont {G.}~\bibnamefont {Baldwin}}, \bibinfo {author}
  {\bibfnamefont {D.~W.}\ \bibnamefont {McComb}}, \bibinfo {author}
  {\bibfnamefont {T.}~\bibnamefont {Albrecht}}, \ and\ \bibinfo {author}
  {\bibfnamefont {J.~B.}\ \bibnamefont {Edel}},\ }\href {\doibase
  10.1021/nl103873a} {\bibfield  {journal} {\bibinfo  {journal} {Nano Letters}\
  }\textbf {\bibinfo {volume} {11}},\ \bibinfo {pages} {279} (\bibinfo {year}
  {2011})},\ \Eprint
  {http://arxiv.org/abs/http://pubs.acs.org/doi/pdf/10.1021/nl103873a}
  {http://pubs.acs.org/doi/pdf/10.1021/nl103873a} \BibitemShut {NoStop}%
\bibitem [{\citenamefont {Han}, \citenamefont {Fu},\ and\ \citenamefont
  {Schoch}(2008)}]{Han:2008}%
  \BibitemOpen
  \bibfield  {author} {\bibinfo {author} {\bibfnamefont {J.}~\bibnamefont
  {Han}}, \bibinfo {author} {\bibfnamefont {J.}~\bibnamefont {Fu}}, \ and\
  \bibinfo {author} {\bibfnamefont {R.~B.}\ \bibnamefont {Schoch}},\ }\href
  {\doibase 10.1039/B714128A} {\bibfield  {journal} {\bibinfo  {journal} {Lab
  Chip}\ }\textbf {\bibinfo {volume} {8}},\ \bibinfo {pages} {23} (\bibinfo
  {year} {2008})}\BibitemShut {NoStop}%
\bibitem [{\citenamefont {Losic}\ and\ \citenamefont
  {Simovic}(2009)}]{Losic:2009}%
  \BibitemOpen
  \bibfield  {author} {\bibinfo {author} {\bibfnamefont {D.}~\bibnamefont
  {Losic}}\ and\ \bibinfo {author} {\bibfnamefont {S.}~\bibnamefont
  {Simovic}},\ }\href {\doibase 10.1517/17425240903300857} {\bibfield
  {journal} {\bibinfo  {journal} {Expert Opin. Drug. Deliv.}\ }\textbf
  {\bibinfo {volume} {6}},\ \bibinfo {pages} {1363} (\bibinfo {year} {2009})},\
  \bibinfo {note} {pMID: 19860534},\ \Eprint
  {http://arxiv.org/abs/http://informahealthcare.com/doi/pdf/10.1517/174252409%
03300857} {http://informahealthcare.com/doi/pdf/10.1517/17425240903300857}
  \BibitemShut {NoStop}%
\bibitem [{\citenamefont {Cherf}\ \emph {et~al.}(2012)\citenamefont {Cherf},
  \citenamefont {Lieberman}, \citenamefont {Rashid}, \citenamefont {Lam},
  \citenamefont {Karplus},\ and\ \citenamefont {Akeson}}]{Cherf:2012}%
  \BibitemOpen
  \bibfield  {author} {\bibinfo {author} {\bibfnamefont {G.~M.}\ \bibnamefont
  {Cherf}}, \bibinfo {author} {\bibfnamefont {K.~R.}\ \bibnamefont
  {Lieberman}}, \bibinfo {author} {\bibfnamefont {H.}~\bibnamefont {Rashid}},
  \bibinfo {author} {\bibfnamefont {C.~E.}\ \bibnamefont {Lam}}, \bibinfo
  {author} {\bibfnamefont {K.}~\bibnamefont {Karplus}}, \ and\ \bibinfo
  {author} {\bibfnamefont {M.}~\bibnamefont {Akeson}},\ }\href {\doibase
  10.1038/nbt.2147} {\bibfield  {journal} {\bibinfo  {journal} {Nat. Biotech.}\
  }\textbf {\bibinfo {volume} {30}},\ \bibinfo {pages} {344} (\bibinfo {year}
  {2012})}\BibitemShut {NoStop}%
\bibitem [{\citenamefont {Manrao}\ \emph {et~al.}(2012)\citenamefont {Manrao},
  \citenamefont {Derrington}, \citenamefont {Laszlo}, \citenamefont {Langford},
  \citenamefont {Hopper}, \citenamefont {Gillgren}, \citenamefont {Pavlenok},
  \citenamefont {Niederweis},\ and\ \citenamefont {Gundlach}}]{Manrao:2012}%
  \BibitemOpen
  \bibfield  {author} {\bibinfo {author} {\bibfnamefont {E.~A.}\ \bibnamefont
  {Manrao}}, \bibinfo {author} {\bibfnamefont {I.~M.}\ \bibnamefont
  {Derrington}}, \bibinfo {author} {\bibfnamefont {A.~H.}\ \bibnamefont
  {Laszlo}}, \bibinfo {author} {\bibfnamefont {K.~W.}\ \bibnamefont
  {Langford}}, \bibinfo {author} {\bibfnamefont {M.~K.}\ \bibnamefont
  {Hopper}}, \bibinfo {author} {\bibfnamefont {N.}~\bibnamefont {Gillgren}},
  \bibinfo {author} {\bibfnamefont {M.}~\bibnamefont {Pavlenok}}, \bibinfo
  {author} {\bibfnamefont {M.}~\bibnamefont {Niederweis}}, \ and\ \bibinfo
  {author} {\bibfnamefont {J.~H.}\ \bibnamefont {Gundlach}},\ }\href {\doibase
  10.1038/nbt.2171} {\bibfield  {journal} {\bibinfo  {journal} {Nat. Biotech.}\
  }\textbf {\bibinfo {volume} {30}},\ \bibinfo {pages} {349} (\bibinfo {year}
  {2012})}\BibitemShut {NoStop}%
\bibitem [{\citenamefont {Luan}, \citenamefont {Stolovitzky},\ and\
  \citenamefont {Martyna}(2012)}]{Luan:2012}%
  \BibitemOpen
  \bibfield  {author} {\bibinfo {author} {\bibfnamefont {B.}~\bibnamefont
  {Luan}}, \bibinfo {author} {\bibfnamefont {G.}~\bibnamefont {Stolovitzky}}, \
  and\ \bibinfo {author} {\bibfnamefont {G.}~\bibnamefont {Martyna}},\ }\href
  {\doibase 10.1039/C1NR11201E} {\bibfield  {journal} {\bibinfo  {journal}
  {Nanoscale}\ }\textbf {\bibinfo {volume} {4}},\ \bibinfo {pages} {1068}
  (\bibinfo {year} {2012})}\BibitemShut {NoStop}%
\bibitem [{\citenamefont {Meller}\ \emph {et~al.}(2000)\citenamefont {Meller},
  \citenamefont {Nivon}, \citenamefont {Brandin}, \citenamefont {Golovchenko},\
  and\ \citenamefont {Branton}}]{Meller:2000}%
  \BibitemOpen
  \bibfield  {author} {\bibinfo {author} {\bibfnamefont {A.}~\bibnamefont
  {Meller}}, \bibinfo {author} {\bibfnamefont {L.}~\bibnamefont {Nivon}},
  \bibinfo {author} {\bibfnamefont {E.}~\bibnamefont {Brandin}}, \bibinfo
  {author} {\bibfnamefont {J.}~\bibnamefont {Golovchenko}}, \ and\ \bibinfo
  {author} {\bibfnamefont {D.}~\bibnamefont {Branton}},\ }\href {\doibase
  10.1073/pnas.97.3.1079} {\bibfield  {journal} {\bibinfo  {journal} {Proc.
  Natl. Acad. Sci. USA}\ }\textbf {\bibinfo {volume} {97}},\ \bibinfo {pages}
  {1079} (\bibinfo {year} {2000})},\ \Eprint
  {http://arxiv.org/abs/http://www.pnas.org/content/97/3/1079.full.pdf+html}
  {http://www.pnas.org/content/97/3/1079.full.pdf+html} \BibitemShut {NoStop}%
\bibitem [{\citenamefont {Meller}, \citenamefont {Nivon},\ and\ \citenamefont
  {Branton}(2001)}]{Meller:2001}%
  \BibitemOpen
  \bibfield  {author} {\bibinfo {author} {\bibfnamefont {A.}~\bibnamefont
  {Meller}}, \bibinfo {author} {\bibfnamefont {L.}~\bibnamefont {Nivon}}, \
  and\ \bibinfo {author} {\bibfnamefont {D.}~\bibnamefont {Branton}},\ }\href
  {\doibase 10.1103/PhysRevLett.86.3435} {\bibfield  {journal} {\bibinfo
  {journal} {Phys. Rev. Lett.}\ }\textbf {\bibinfo {volume} {86}},\ \bibinfo
  {pages} {3435} (\bibinfo {year} {2001})}\BibitemShut {NoStop}%
\bibitem [{\citenamefont {Li}\ and\ \citenamefont {Talaga}(2010)}]{Li:2010}%
  \BibitemOpen
  \bibfield  {author} {\bibinfo {author} {\bibfnamefont {J.}~\bibnamefont
  {Li}}\ and\ \bibinfo {author} {\bibfnamefont {D.~S.}\ \bibnamefont
  {Talaga}},\ }\href {http://stacks.iop.org/0953-8984/22/i=45/a=454129}
  {\bibfield  {journal} {\bibinfo  {journal} {J. Phys.: Condens. Matter}\
  }\textbf {\bibinfo {volume} {22}},\ \bibinfo {pages} {454129} (\bibinfo
  {year} {2010})}\BibitemShut {NoStop}%
\bibitem [{\citenamefont {Kasianowicz}\ \emph {et~al.}(1996)\citenamefont
  {Kasianowicz}, \citenamefont {Brandin}, \citenamefont {Branton},\ and\
  \citenamefont {Deamer}}]{Kasianowicz:1996}%
  \BibitemOpen
  \bibfield  {author} {\bibinfo {author} {\bibfnamefont {J.}~\bibnamefont
  {Kasianowicz}}, \bibinfo {author} {\bibfnamefont {E.}~\bibnamefont
  {Brandin}}, \bibinfo {author} {\bibfnamefont {D.}~\bibnamefont {Branton}}, \
  and\ \bibinfo {author} {\bibfnamefont {D.}~\bibnamefont {Deamer}},\ }\href
  {http://www.pnas.org/content/93/24/13770.abstract} {\bibfield  {journal}
  {\bibinfo  {journal} {Proc. Natl. Acad. Sci. USA}\ }\textbf {\bibinfo
  {volume} {93}},\ \bibinfo {pages} {13770} (\bibinfo {year} {1996})},\ \Eprint
  {http://arxiv.org/abs/http://www.pnas.org/content/93/24/13770.full.pdf+html}
  {http://www.pnas.org/content/93/24/13770.full.pdf+html} \BibitemShut
  {NoStop}%
\bibitem [{\citenamefont {Hall}\ \emph {et~al.}(2010)\citenamefont {Hall},
  \citenamefont {Scott}, \citenamefont {Rotem}, \citenamefont {Mehta},
  \citenamefont {Bayley},\ and\ \citenamefont {Dekker}}]{Hall:2010}%
  \BibitemOpen
  \bibfield  {author} {\bibinfo {author} {\bibfnamefont {A.~R.}\ \bibnamefont
  {Hall}}, \bibinfo {author} {\bibfnamefont {A.}~\bibnamefont {Scott}},
  \bibinfo {author} {\bibfnamefont {D.}~\bibnamefont {Rotem}}, \bibinfo
  {author} {\bibfnamefont {K.~K.}\ \bibnamefont {Mehta}}, \bibinfo {author}
  {\bibfnamefont {H.}~\bibnamefont {Bayley}}, \ and\ \bibinfo {author}
  {\bibfnamefont {C.}~\bibnamefont {Dekker}},\ }\href {\doibase
  http://dx.doi.org/10.1038/nnano.2010.237} {\bibfield  {journal} {\bibinfo
  {journal} {Nat. Nanotechnol.}\ }\textbf {\bibinfo {volume} {5}},\ \bibinfo
  {pages} {874} (\bibinfo {year} {2010})}\BibitemShut {NoStop}%
\bibitem [{\citenamefont {Wanunu}\ and\ \citenamefont
  {Meller}(2007)}]{Wanunu:2007}%
  \BibitemOpen
  \bibfield  {author} {\bibinfo {author} {\bibfnamefont {M.}~\bibnamefont
  {Wanunu}}\ and\ \bibinfo {author} {\bibfnamefont {A.}~\bibnamefont
  {Meller}},\ }\href {\doibase 10.1021/nl070462b} {\bibfield  {journal}
  {\bibinfo  {journal} {Nano Lett.}\ }\textbf {\bibinfo {volume} {7}},\
  \bibinfo {pages} {1580} (\bibinfo {year} {2007})}\BibitemShut {NoStop}%
\bibitem [{\citenamefont {Kowalczyk}\ \emph {et~al.}(2011)\citenamefont
  {Kowalczyk}, \citenamefont {Kapinos}, \citenamefont {Blosser}, \citenamefont
  {Magalhaes}, \citenamefont {van Nies}, \citenamefont {Lim},\ and\
  \citenamefont {Dekker}}]{Kowalczyk:2011}%
  \BibitemOpen
  \bibfield  {author} {\bibinfo {author} {\bibfnamefont {S.~W.}\ \bibnamefont
  {Kowalczyk}}, \bibinfo {author} {\bibfnamefont {L.}~\bibnamefont {Kapinos}},
  \bibinfo {author} {\bibfnamefont {T.~R.}\ \bibnamefont {Blosser}}, \bibinfo
  {author} {\bibfnamefont {T.}~\bibnamefont {Magalhaes}}, \bibinfo {author}
  {\bibfnamefont {P.}~\bibnamefont {van Nies}}, \bibinfo {author}
  {\bibfnamefont {R.~Y.~H.}\ \bibnamefont {Lim}}, \ and\ \bibinfo {author}
  {\bibfnamefont {C.}~\bibnamefont {Dekker}},\ }\href {\doibase DOI:
  10.1038/nnano.2011.88} {\bibfield  {journal} {\bibinfo  {journal} {Nat.
  Nanotechnol.}\ }\textbf {\bibinfo {volume} {6}},\ \bibinfo {pages} {433 }
  (\bibinfo {year} {2011})}\BibitemShut {NoStop}%
\bibitem [{\citenamefont {Wei}\ \emph {et~al.}(2012)\citenamefont {Wei},
  \citenamefont {Gatterdam}, \citenamefont {Wieneke}, \citenamefont {Tampe},\
  and\ \citenamefont {Rant}}]{Wei:2012}%
  \BibitemOpen
  \bibfield  {author} {\bibinfo {author} {\bibfnamefont {R.}~\bibnamefont
  {Wei}}, \bibinfo {author} {\bibfnamefont {V.}~\bibnamefont {Gatterdam}},
  \bibinfo {author} {\bibfnamefont {R.}~\bibnamefont {Wieneke}}, \bibinfo
  {author} {\bibfnamefont {R.}~\bibnamefont {Tampe}}, \ and\ \bibinfo {author}
  {\bibfnamefont {U.}~\bibnamefont {Rant}},\ }\href {\doibase
  10.1038/nnano.2012.24} {\bibfield  {journal} {\bibinfo  {journal} {Nat.
  Nanotechnol.}\ }\textbf {\bibinfo {volume} {7}},\ \bibinfo {pages} {257}
  (\bibinfo {year} {2012})}\BibitemShut {NoStop}%
\bibitem [{\citenamefont {Hou}, \citenamefont {Guo},\ and\ \citenamefont
  {Jiang}(2011)}]{Hou:2011}%
  \BibitemOpen
  \bibfield  {author} {\bibinfo {author} {\bibfnamefont {X.}~\bibnamefont
  {Hou}}, \bibinfo {author} {\bibfnamefont {W.}~\bibnamefont {Guo}}, \ and\
  \bibinfo {author} {\bibfnamefont {L.}~\bibnamefont {Jiang}},\ }\href
  {\doibase 10.1039/C0CS00053A} {\bibfield  {journal} {\bibinfo  {journal}
  {Chem. Soc. Rev.}\ }\textbf {\bibinfo {volume} {40}},\ \bibinfo {pages}
  {2385} (\bibinfo {year} {2011})}\BibitemShut {NoStop}%
\bibitem [{\citenamefont {Adiga}\ and\ \citenamefont
  {Brenner}(2005)}]{Adiga:2005}%
  \BibitemOpen
  \bibfield  {author} {\bibinfo {author} {\bibfnamefont {S.~P.}\ \bibnamefont
  {Adiga}}\ and\ \bibinfo {author} {\bibfnamefont {D.~W.}\ \bibnamefont
  {Brenner}},\ }\href {\doibase 10.1021/nl051843x} {\bibfield  {journal}
  {\bibinfo  {journal} {Nano Lett.}\ }\textbf {\bibinfo {volume} {5}},\
  \bibinfo {pages} {2509} (\bibinfo {year} {2005})}\BibitemShut {NoStop}%
\bibitem [{\citenamefont {Adiga}\ and\ \citenamefont
  {Brenner}(2012)}]{Adiga:2012}%
  \BibitemOpen
  \bibfield  {author} {\bibinfo {author} {\bibfnamefont {S.~P.}\ \bibnamefont
  {Adiga}}\ and\ \bibinfo {author} {\bibfnamefont {D.~W.}\ \bibnamefont
  {Brenner}},\ }\href@noop {} {\bibfield  {journal} {\bibinfo  {journal} {J.
  Funct. Biomat.}\ }\textbf {\bibinfo {volume} {3}},\ \bibinfo {pages} {239}
  (\bibinfo {year} {2012})}\BibitemShut {NoStop}%
\bibitem [{\citenamefont {Hoelz}, \citenamefont {Debler},\ and\ \citenamefont
  {Blobel}(2011)}]{Hoelz:2011}%
  \BibitemOpen
  \bibfield  {author} {\bibinfo {author} {\bibfnamefont {A.}~\bibnamefont
  {Hoelz}}, \bibinfo {author} {\bibfnamefont {E.~W.}\ \bibnamefont {Debler}}, \
  and\ \bibinfo {author} {\bibfnamefont {G.}~\bibnamefont {Blobel}},\ }\href
  {\doibase 10.1146/annurev-biochem-060109-151030} {\bibfield  {journal}
  {\bibinfo  {journal} {Annu. Rev. Biochem.}\ }\textbf {\bibinfo {volume}
  {80}},\ \bibinfo {pages} {613} (\bibinfo {year} {2011})}\BibitemShut
  {NoStop}%
\bibitem [{\citenamefont {Grossman}, \citenamefont {Medalia},\ and\
  \citenamefont {Zwerger}(2012)}]{Grossman:2012}%
  \BibitemOpen
  \bibfield  {author} {\bibinfo {author} {\bibfnamefont {E.}~\bibnamefont
  {Grossman}}, \bibinfo {author} {\bibfnamefont {O.}~\bibnamefont {Medalia}}, \
  and\ \bibinfo {author} {\bibfnamefont {M.}~\bibnamefont {Zwerger}},\ }\href
  {\doibase 10.1146/annurev-biophys-050511-102328} {\bibfield  {journal}
  {\bibinfo  {journal} {Annual Review of Biophysics}\ }\textbf {\bibinfo
  {volume} {41}},\ \bibinfo {pages} {557} (\bibinfo {year} {2012})},\ \bibinfo
  {note} {pMID: 22577827},\ \Eprint
  {http://arxiv.org/abs/http://www.annualreviews.org/doi/pdf/10.1146/annurev-b%
iophys-050511-102328}
  {http://www.annualreviews.org/doi/pdf/10.1146/annurev-biophys-050511-102328}
  \BibitemShut {NoStop}%
\bibitem [{\citenamefont {Jovanovic-Talisman}\ \emph
  {et~al.}(2009)\citenamefont {Jovanovic-Talisman}, \citenamefont
  {Tetenbaum-Novatt}, \citenamefont {McKenney}, \citenamefont {Zilman},
  \citenamefont {Peters}, \citenamefont {Rout},\ and\ \citenamefont
  {Chait}}]{Jovanovic:2009}%
  \BibitemOpen
  \bibfield  {author} {\bibinfo {author} {\bibfnamefont {T.}~\bibnamefont
  {Jovanovic-Talisman}}, \bibinfo {author} {\bibfnamefont {J.}~\bibnamefont
  {Tetenbaum-Novatt}}, \bibinfo {author} {\bibfnamefont {A.~S.}\ \bibnamefont
  {McKenney}}, \bibinfo {author} {\bibfnamefont {A.}~\bibnamefont {Zilman}},
  \bibinfo {author} {\bibfnamefont {R.}~\bibnamefont {Peters}}, \bibinfo
  {author} {\bibfnamefont {M.~P.}\ \bibnamefont {Rout}}, \ and\ \bibinfo
  {author} {\bibfnamefont {B.~T.}\ \bibnamefont {Chait}},\ }\href {\doibase
  10.1038/nature07600} {\bibfield  {journal} {\bibinfo  {journal} {Nature}\
  }\textbf {\bibinfo {volume} {457}},\ \bibinfo {pages} {1023} (\bibinfo {year}
  {2009})}\BibitemShut {NoStop}%
\bibitem [{\citenamefont {Kowalczyk}, \citenamefont {Blosser},\ and\
  \citenamefont {Dekker}(2011)}]{Kowalczyk:2011:2}%
  \BibitemOpen
  \bibfield  {author} {\bibinfo {author} {\bibfnamefont {S.~W.}\ \bibnamefont
  {Kowalczyk}}, \bibinfo {author} {\bibfnamefont {T.~R.}\ \bibnamefont
  {Blosser}}, \ and\ \bibinfo {author} {\bibfnamefont {C.}~\bibnamefont
  {Dekker}},\ }\href {\doibase 10.1016/j.tibtech.2011.07.006} {\bibfield
  {journal} {\bibinfo  {journal} {Trends Biotechnol.}\ }\textbf {\bibinfo
  {volume} {29}},\ \bibinfo {pages} {607 } (\bibinfo {year}
  {2011})}\BibitemShut {NoStop}%
\bibitem [{\citenamefont {Peleg}\ \emph {et~al.}(2011)\citenamefont {Peleg},
  \citenamefont {Tagliazucchi}, \citenamefont {Kr{{\"o}}ger}, \citenamefont
  {Rabin},\ and\ \citenamefont {Szleifer}}]{Peleg:2011}%
  \BibitemOpen
  \bibfield  {author} {\bibinfo {author} {\bibfnamefont {O.}~\bibnamefont
  {Peleg}}, \bibinfo {author} {\bibfnamefont {M.}~\bibnamefont {Tagliazucchi}},
  \bibinfo {author} {\bibfnamefont {M.}~\bibnamefont {Kr{{\"o}}ger}}, \bibinfo
  {author} {\bibfnamefont {Y.}~\bibnamefont {Rabin}}, \ and\ \bibinfo {author}
  {\bibfnamefont {I.}~\bibnamefont {Szleifer}},\ }\href {\doibase
  10.1021/nn200702u} {\bibfield  {journal} {\bibinfo  {journal} {ACS Nano}\
  }\textbf {\bibinfo {volume} {5}},\ \bibinfo {pages} {4737} (\bibinfo {year}
  {2011})}\BibitemShut {NoStop}%
\bibitem [{\citenamefont {Moussavi-Baygi}\ \emph {et~al.}(2011)\citenamefont
  {Moussavi-Baygi}, \citenamefont {Jamali}, \citenamefont {Karimi},\ and\
  \citenamefont {Mofrad}}]{Moussavi:2011}%
  \BibitemOpen
  \bibfield  {author} {\bibinfo {author} {\bibfnamefont {R.}~\bibnamefont
  {Moussavi-Baygi}}, \bibinfo {author} {\bibfnamefont {Y.}~\bibnamefont
  {Jamali}}, \bibinfo {author} {\bibfnamefont {R.}~\bibnamefont {Karimi}}, \
  and\ \bibinfo {author} {\bibfnamefont {M.~R.~K.}\ \bibnamefont {Mofrad}},\
  }\href {\doibase 10.1016/j.bpj.2011.01.061} {\bibfield  {journal} {\bibinfo
  {journal} {Biophys. J.}\ }\textbf {\bibinfo {volume} {100}},\ \bibinfo
  {pages} {1410} (\bibinfo {year} {2011})}\BibitemShut {NoStop}%
\bibitem [{\citenamefont {Mincer}\ and\ \citenamefont
  {Simon}(2011)}]{Mincer:2011}%
  \BibitemOpen
  \bibfield  {author} {\bibinfo {author} {\bibfnamefont {J.~S.}\ \bibnamefont
  {Mincer}}\ and\ \bibinfo {author} {\bibfnamefont {S.~M.}\ \bibnamefont
  {Simon}},\ }\href {\doibase 10.1073/pnas.1104521108} {\bibfield  {journal}
  {\bibinfo  {journal} {Proc. Natl. Acad. Sci. USA}\ }\textbf {\bibinfo
  {volume} {108}},\ \bibinfo {pages} {E351} (\bibinfo {year}
  {2011})}\BibitemShut {NoStop}%
\bibitem [{\citenamefont {Osmanovic}\ \emph {et~al.}(2012)\citenamefont
  {Osmanovic}, \citenamefont {Bailey}, \citenamefont {Harker}, \citenamefont
  {Fassati}, \citenamefont {Hoogenboom},\ and\ \citenamefont
  {Ford}}]{Osmanovic:2012}%
  \BibitemOpen
  \bibfield  {author} {\bibinfo {author} {\bibfnamefont {D.}~\bibnamefont
  {Osmanovic}}, \bibinfo {author} {\bibfnamefont {J.}~\bibnamefont {Bailey}},
  \bibinfo {author} {\bibfnamefont {A.~H.}\ \bibnamefont {Harker}}, \bibinfo
  {author} {\bibfnamefont {A.}~\bibnamefont {Fassati}}, \bibinfo {author}
  {\bibfnamefont {B.~W.}\ \bibnamefont {Hoogenboom}}, \ and\ \bibinfo {author}
  {\bibfnamefont {I.~J.}\ \bibnamefont {Ford}},\ }\href@noop {} {\bibfield
  {journal} {\bibinfo  {journal} {Phys. Rev. E}\ }\textbf {\bibinfo {volume}
  {85}},\ \bibinfo {pages} {061917} (\bibinfo {year} {2012})}\BibitemShut
  {NoStop}%
\bibitem [{\citenamefont {Egorov}\ \emph {et~al.}(2011)\citenamefont {Egorov},
  \citenamefont {Milchev}, \citenamefont {Klushin},\ and\ \citenamefont
  {Binder}}]{Egorov:2011}%
  \BibitemOpen
  \bibfield  {author} {\bibinfo {author} {\bibfnamefont {S.~A.}\ \bibnamefont
  {Egorov}}, \bibinfo {author} {\bibfnamefont {A.}~\bibnamefont {Milchev}},
  \bibinfo {author} {\bibfnamefont {L.}~\bibnamefont {Klushin}}, \ and\
  \bibinfo {author} {\bibfnamefont {K.}~\bibnamefont {Binder}},\ }\href
  {\doibase 10.1039/C1SM05139C} {\bibfield  {journal} {\bibinfo  {journal}
  {Soft Matter}\ }\textbf {\bibinfo {volume} {7}},\ \bibinfo {pages} {5669}
  (\bibinfo {year} {2011})}\BibitemShut {NoStop}%
\bibitem [{\citenamefont {Kreer}\ \emph {et~al.}(2004)\citenamefont {Kreer},
  \citenamefont {Metzger}, \citenamefont {M{\"{u}}ller}, \citenamefont
  {Binder},\ and\ \citenamefont {Baschnagel}}]{Kreer:2004}%
  \BibitemOpen
  \bibfield  {author} {\bibinfo {author} {\bibfnamefont {T.}~\bibnamefont
  {Kreer}}, \bibinfo {author} {\bibfnamefont {S.}~\bibnamefont {Metzger}},
  \bibinfo {author} {\bibfnamefont {M.}~\bibnamefont {M{\"{u}}ller}}, \bibinfo
  {author} {\bibfnamefont {K.}~\bibnamefont {Binder}}, \ and\ \bibinfo {author}
  {\bibfnamefont {J.}~\bibnamefont {Baschnagel}},\ }\href@noop {} {\bibfield
  {journal} {\bibinfo  {journal} {J. Chem. Phys.}\ }\textbf {\bibinfo {volume}
  {120}} (\bibinfo {year} {2004})}\BibitemShut {NoStop}%
\bibitem [{\citenamefont {Osmanovi{\'{c}}}\ \emph {et~al.}(2013)\citenamefont
  {Osmanovi{\'{c}}}, \citenamefont {Fassati}, \citenamefont {Ford},\ and\
  \citenamefont {Hoogenboom}}]{Osmanovic:2013}%
  \BibitemOpen
  \bibfield  {author} {\bibinfo {author} {\bibfnamefont {D.}~\bibnamefont
  {Osmanovi{\'{c}}}}, \bibinfo {author} {\bibfnamefont {A.}~\bibnamefont
  {Fassati}}, \bibinfo {author} {\bibfnamefont {I.~J.}\ \bibnamefont {Ford}}, \
  and\ \bibinfo {author} {\bibfnamefont {B.~W.}\ \bibnamefont {Hoogenboom}},\
  }\href {\doibase 10.1039/C3SM50722J} {\bibfield  {journal} {\bibinfo
  {journal} {Soft Matter}\ }\textbf {\bibinfo {volume} {9}},\ \bibinfo {pages}
  {10442} (\bibinfo {year} {2013})}\BibitemShut {NoStop}%
\bibitem [{\citenamefont {Zaitseva}, \citenamefont {Myers},\ and\ \citenamefont
  {Fassati}(2006)}]{Zaitseva:2006}%
  \BibitemOpen
  \bibfield  {author} {\bibinfo {author} {\bibfnamefont {L.}~\bibnamefont
  {Zaitseva}}, \bibinfo {author} {\bibfnamefont {R.}~\bibnamefont {Myers}}, \
  and\ \bibinfo {author} {\bibfnamefont {A.}~\bibnamefont {Fassati}},\ }\href
  {\doibase 10.1371/journal.pbio.0040332} {\bibfield  {journal} {\bibinfo
  {journal} {PLoS Biol}\ }\textbf {\bibinfo {volume} {4}},\ \bibinfo {pages}
  {e332} (\bibinfo {year} {2006})}\BibitemShut {NoStop}%
\bibitem [{\citenamefont {Metropolis}(1953)}]{Metropolis:1953}%
  \BibitemOpen
  \bibfield  {author} {\bibinfo {author} {\bibfnamefont {N.}~\bibnamefont
  {Metropolis}},\ }\href@noop {} {\bibfield  {journal} {\bibinfo  {journal} {J.
  Chem. Phys.}\ }\textbf {\bibinfo {volume} {21}} (\bibinfo {year}
  {1953})}\BibitemShut {NoStop}%
\bibitem [{\citenamefont {Mulero}\ \emph {et~al.}(2008)\citenamefont {Mulero},
  \citenamefont {Galán}, \citenamefont {Parra},\ and\ \citenamefont
  {Cuadros}}]{Mulero:20082}%
  \BibitemOpen
  \bibfield  {author} {\bibinfo {author} {\bibfnamefont {A.}~\bibnamefont
  {Mulero}}, \bibinfo {author} {\bibfnamefont {C.}~\bibnamefont {Galán}},
  \bibinfo {author} {\bibfnamefont {M.}~\bibnamefont {Parra}}, \ and\ \bibinfo
  {author} {\bibfnamefont {F.}~\bibnamefont {Cuadros}},\ }in\ \href {\doibase
  10.1007/978-3-540-78767-9_3} {\emph {\bibinfo {booktitle} {Theory and
  Simulation of Hard-Sphere Fluids and Related Systems}}},\ \bibinfo {series}
  {Lecture Notes in Physics}, Vol.\ \bibinfo {volume} {753},\ \bibinfo {editor}
  {edited by\ \bibinfo {editor} {\bibfnamefont {A.}~\bibnamefont {Mulero}}}\
  (\bibinfo  {publisher} {Springer Berlin Heidelberg},\ \bibinfo {year}
  {2008})\ pp.\ \bibinfo {pages} {37--109}\BibitemShut {NoStop}%
\bibitem [{\citenamefont {Wertheim}(1987)}]{Wertheim:1987}%
  \BibitemOpen
  \bibfield  {author} {\bibinfo {author} {\bibfnamefont {M.~S.}\ \bibnamefont
  {Wertheim}},\ }\href {\doibase 10.1063/1.453326} {\bibfield  {journal}
  {\bibinfo  {journal} {J. Chem. Phys.}\ }\textbf {\bibinfo {volume} {87}},\
  \bibinfo {pages} {7323} (\bibinfo {year} {1987})}\BibitemShut {NoStop}%
\bibitem [{\citenamefont {Egorov}(2008)}]{Egorov:2008}%
  \BibitemOpen
  \bibfield  {author} {\bibinfo {author} {\bibfnamefont {S.~A.}\ \bibnamefont
  {Egorov}},\ }\href {\doibase 10.1063/1.2968545} {\bibfield  {journal}
  {\bibinfo  {journal} {J. Chem. Phys.}\ }\textbf {\bibinfo {volume} {129}},\
  \bibinfo {eid} {064901} (\bibinfo {year} {2008})}\BibitemShut {NoStop}%
\end{thebibliography}
%

\end{document}